\documentclass[11pt]{article}
\usepackage{amsmath}
\usepackage{mathrsfs,epsfig,morefloats}
\usepackage{amsfonts}
\usepackage{mathtools}
\usepackage{authblk}
\usepackage{xcolor}
\usepackage{graphicx}
\usepackage{float}
\usepackage{amsthm}
\usepackage{amssymb}
\usepackage{lscape}
\usepackage{rotating}
\usepackage{subfigure}
\usepackage{graphicx}

\usepackage{tabularx}
\usepackage{multirow}

\usepackage{algorithm}

\allowdisplaybreaks

\usepackage{algpseudocode}

\usepackage{caption}
\usepackage{lipsum}

\DeclareCaptionFormat{algor}{%
  \hrulefill\par\offinterlineskip\vskip1pt%
    \textbf{#1#2}#3\offinterlineskip\hrulefill}
\DeclareCaptionStyle{algori}{singlelinecheck=off,format=algor,labelsep=space}
\usepackage{hyperref}
\hypersetup{
    colorlinks=true,
    linkcolor=blue,
    filecolor=blue,      
    urlcolor=blue,
citecolor=blue,
}
 \usepackage[sort&compress]{natbib}

\usepackage[section]{placeins}
\newcommand{\Date}[1]{\def\@Date{#1}}
\def\today{\number\day~\ifcase\month\or
 January\or February\or March\or April\or May\or June\or
 July\or August\or September\or October\or November\or December\fi~\number\year}

\def\be{\begin{equation}}
\def\ee{\end{equation}}
\def\bea{\begin{eqnarray}}
\def\eea{\end{eqnarray}}
\def\bd{\begin{displaymath}}
\def\ed{\end{displaymath}}
\def\bda{\begin{eqnarray*}}
\def\eda{\end{eqnarray*}}
\def\bsm{\begin{small}}
\def\esm{\end{small}}

\def\t0{\theta_0}

\def\ha1{\hat \beta_1}

\def\bnt{\begin{enumerate}}
\def\ent{\end{enumerate}}

\def\bsc{\begin{scriptsize}}
\def\esc{\end{scriptsize}}

\theoremstyle{definition}

\newcommand{\E}{\mathbb{E}}

\makeatletter
\newcommand{\figcaption}{\def\@captype{figure}\caption}
\newcommand{\tabcaption}{\def\@captype{table}\caption}
\makeatother

\usepackage{scalerel,stackengine}
\stackMath
\newcommand\reallywidehat[1]{%
\savestack{\tmpbox}{\stretchto{%
  \scaleto{%
    \scalerel*[\widthof{\ensuremath{#1}}]{\kern-.6pt\bigwedge\kern-.6pt}%
    {\rule[-\textheight/2]{1ex}{\textheight}}
  }{\textheight}%
}{0.5ex}}%
\stackon[1pt]{#1}{\tmpbox}%
}
\makeatletter
\newcommand{\printfnsymbol}[1]{%
  \textsuperscript{\@fnsymbol{#1}}%
}
\makeatother

\usepackage{bbm}
\usepackage{tikz}
\usetikzlibrary{arrows.meta}

\usepackage{authblk}

\title{A Zero-Inflated Beta Mixture Model for Marginal Mediation Analysis with Compositional Microbiome Mediators}

\author{Seungjun Ahn$^{1,*}$, Quran Wu$^{2,*}$, Alicia Yang$^{3}$, and Zhigang Li$^{2}$}

\affil{$^{1}$Department of Population Health Science and Policy, Icahn School of Medicine at Mount Sinai, New York, NY 10029, U.S.A.}
\affil{$^{2}$Department of Biostatistics, University of Florida, Gainesville, FL 32611, U.S.A.}
\affil{$^{3}$Department of Otolaryngology - Head and Neck Surgery, Icahn School of Medicine at Mount Sinai, New York, NY 10029, U.S.A.}
\affil{$^{*}$These authors contributed equally.}

\date{}

\begin{document}

\maketitle

\begin{abstract}
The role of the microbiome in disease pathogenesis is an emerging field with strong evidence suggesting that dysbiosis is associated with precancerous and cancerous states. Microbiome data present substantial challenges for causal mediation analysis due to sparsity, compositional constraints, and latent heterogeneity. To address these issues, we propose a zero-inflated beta mixture (ZIBM) method for mediation analysis with compositional microbiome mediators. The proposed method accommodates excess zeros through a zero-inflation component and captures heterogeneity in non-zero relative abundances using a beta mixture distribution. Within the potential-outcomes framework, the ZIBM provides estimates of marginal microbiome-mediated causal effects, and model parameters are estimated using an expectation-maximization algorithm. Simulation studies demonstrate that the ZIBM yields more accurate estimation and reliable inference under conditions commonly observed in microbiome data, compared with existing approaches. An application to a real microbiome study further illustrates its practical utility. These results indicate that the proposed method provides a more flexible and robust statistical framework for mediation analysis involving compositional microbiome data.
\end{abstract}


\section{Introduction}
Advances in high-throughput sequencing technologies, including 16S ribosomal RNA gene sequencing and metagenomic shotgun sequencing, have enabled comprehensive characterization of microbial communities. These technologies have facilitated investigations into microbiome-mediated mechanisms in complex diseases such as Alzheimer’s disease and cancer \citep{Wang2019, Jin2019, Tanoue2019, kwak, bautista}. Recent research has focused on the potential role of microbiota in cancer development, particularly for head and neck squamous cell carcinoma and colorectal cancer (CRC) \citep{kwak, bautista}. Oral microbiome dysbiosis had been associated with environmental exposures, oral hygiene, and periodontal disease, as well as cancer \citep{mousavi, kwak}. Similarly, there is research that shows that gut microbiome dysbiosis associated with CRC is characterized by enrichment of pathobionts and the loss of protective commensal bacterial species \citep{bautista}. Changes in the microbiome may be linked with reprogrammed immune, metabolic, neural, and endocrine pathways. These changes in the relative abundance of bacterial populations may be associated with nutrition, air pollution, endocrine disrupting chemicals, and other environmental exposures \citep{mousavi}. Therefore, there is growing interest in identifying microbial taxa that mediate associations between risk factors and health outcomes. 

Mediation analysis provides a statistical framework for quantifying causal pathways involving intermediate variables. It decomposes the total effect of an exposure into direct and indirect components and has been applied across epidemiology, environmental health sciences, and biomedical research. Within the potential-outcomes framework, mediation analysis allows for nonlinear relationships and interaction effects \citep{VanderWeele2009, Imai, VanderWeele2015}. Comprehensive reviews of mediation methodologies and their assumptions are available in the literature \citep{MacKinnon2007, Vander16, Lange2017}. Despite these advances, extending mediation analysis to microbiome studies remains challenging due to the unique distributional characteristics of microbiome sequencing data.

Microbiome data are inherently compositional, sparse, and high-dimensional, with a substantial proportion of zero observations \citep{Li2018}. These zeros may arise from true biological absence or from technical artifacts such as undersampling and detection limits. The coexistence of compositional constraints and zero inflation complicates statistical modeling and inference, particularly in mediation settings. Although several approaches have been proposed to analyze the microbiome as a high-dimensional mediator \citep{Sohn2019, Wang2020, Zhang2019}, effectively accommodating zero-inflated compositional mediators remains an open methodological problem. To address these challenges, the MarZIC framework \citep{MarZIC} was previously developed to model zero-inflated compositional mediators under the potential-outcomes paradigm. MarZIC enables decomposition of mediation effects into components attributable to binary presence-absence changes and continuous abundance variations, thereby providing interpretable insights into microbiome-mediated mechanisms. Due to the compositional structure of microbiome data, the RA of each taxon is always linearly correlated with other taxa, and thus it is often difficult to define and describe the mediation effect of a taxon. Therefore, MarZIC proposed a definition of the marginal mediation effect by fitting each taxon separately one by one. Nevertheless, the distributional heterogeneity of microbiome relative abundance (RA) data necessitates more flexible modeling strategies, such as those based on mixture distributions.

In this work, we propose a Zero-Inflated Beta Mixture (ZIBM) model for marginal mediation analysis with compositional microbiome mediators. The proposed model extends the MarZIC framework by incorporating a mixture-based representation that captures excess zeros and latent heterogeneity in microbial compositions. Using a zero-inflated beta mixture formulation, the proposed model improves the modeling of sparse and compositional mediators while preserving interpretability within the potential-outcomes framework. This flexibility leads to more accurate estimation and inference of microbiome-mediated causal effects. The remainder of this article is organized as follows. Section \ref{sc:model} introduces the proposed model and notation. Sections \ref{sc:estimation} and \ref{sc:EM} present estimation and inference procedures. Section \ref{sc:simulations} evaluates the performance of the proposed method through simulation studies. Section \ref{sc:realdata} demonstrates its utility using a real microbiome data application. Section \ref{sc:discussion} concludes with a discussion.

\section{Notation and Model Specifications}\label{sc:model}
For simplicity, the subject index is omitted from all notations throughout this paper. We will let $Y$, $M$, and $X$ denote the continuous outcome variable, the mediator (i.e., a taxon/ASV/OTU), and the exposure variable, respectively. The outcome $Y$ is assumed to depend on $M$ and $X$ through the following marginal regression model \citep{MarZIC}:
\begin{align}
Y=\beta_0+\beta_1M+\beta_2 \mathbbm{1}_{(M>0)}+\beta_3X+\beta_4X\mathbbm{1}_{(M>0)}+\beta_5XM+\epsilon, \label{ge:1}
\end{align}
where $\mathbbm{1}_{(\cdot)}$ denotes an indicator function, the random error $\epsilon$ follows a normal distribution $N(0,\delta)$, and $\beta_1$, $\beta_2$, $\beta_3$, $\beta_4$, and $\beta_5$ are regression coefficients. The mediator $M$ represents the RA of a microbial taxon, an ASV or an OTU.

The mediation framework involves two pathways arising from the association between $M$ and $X$. Under a popular choice of Dirichlet distribution for the composition of microbial communities, the marginal distribution of a taxon is Beta distribution. Here, we extend it to a mixture of Beta distributions to accommodate complexities in real-word data. The $M$ is modeled using a zero-inflated Beta mixture (ZIBM) distribution, whose two-part density function is given by
\begin{align}
  f(m;\mathbf{\theta})=\begin{cases}
    \Delta, & m=0\\
    (1-\Delta)\Bigl(\sum_{k=1}^K \psi_k \mathrm{Beta}(m \mid \mu_k,\phi) + \Bigl(1-\sum_{k=1}^K \psi_k\Bigr)\mathrm{Beta}(m \mid \mu_{K+1},\phi)\Bigr), &m>0,
  \end{cases} \label{ge:2}
\end{align}
where $\Delta$ denotes the probability of observing a zero. The function $\mathrm{Beta}(m \mid \mu_{k},\phi) = \frac{m^{\mu_{k}\phi-1}(1-m)^{(1-\mu_{k})\phi-1}}{B\big(\mu_{k}\phi,(1-\mu_{k})\phi\big)}$ is the Beta density of the $k$th component of the mixture, $\psi_k$ is the weight of the $k$th component, $B(\cdot,\cdot)$ is the Beta function, and $\mu_{k}$ and $\phi$ represent the mean and dispersion parameters, respectively. The relationship between the mediator and the independent variable is modeled using the following regression models:
\begin{align}
&\ln{\Big(\frac{\mu_k}{1-\mu_k}\Big)}=\alpha_{0k}+\alpha_{1k}X, \hspace{0.2cm} k=1,\dots,K+1 \label{ge:21}\\
&\ln\bigg(\frac{\Delta}{1-\Delta}\bigg)=\gamma_0+\gamma_1X. \label{ge:23}
\end{align}

\noindent Equations (\ref{ge:1}) to (\ref{ge:23}) jointly define the full mediation model of the proposed ZIBM approach.

\subsection{Mechanism for Observing Zeros of the Mediator}\label{sc:mechanismzero}
For microbiome abundance data, observations that cannot be detected are set to be zero. Consequently, the observed abundance data may contain two types of zeros: true zeros representing taxon absence and false zeros arising from measurement (or technical) failure. Let $M$ denote the RA of a microbial taxon and let $M^*$ represent its observed value. When the observed value is positive (i.e., $M^*>0$), we assume that $M=M^*$. However, when $M^*=0$, it is not clear whether the true abundance is zero or positive but recorded as zero. In this study, we consider the following mechanism for observing zeros in microbial taxon abundance with these zero characteristics:
\begin{align}\label{eq:RAzero1}
\Pr(M^*=0 \mid M,L)=\mathbbm{1}_{(ML < 1)},
\end{align}
where $L$ is the library size (i.e., sequencing depth) and the product $ML$ can be interpreted as the sample absolute abundance (SAA) of a taxon in a sample. We refer to this mechanism as ``LOD mechanism'' since all SAA values below 1, limit of detection (LOD), cannot be detected.

\subsection{Marginal Mediation Effect and Direct Effect}\label{NIE_NDE}
Within the potential-outcomes (PO) framework \citep{Vander16}, we define the natural indirect effect (NIE) and natural direct effect (NDE), where the NIE represents the marginal mediation effect \citep{MarZIC}. The total effect of the exposure variable $X$ can be expressed as the sum of the NIE and NDE. Let $M_x$ denote the value of $M$ when $X=x$, and let $Y_{xm}$ denote the value of $Y$ when $(X,M)=(x,m)$. The average NIE and NDE corresponding to a change in $X$ from $x_1$ to $x_2$ are defined as follows:
\begin{align*}
   \text{NIE}&=\E\bigl[Y_{x_2M_{x_2}}-Y_{x_2M_{x_1}}\bigr], \\
   \text{NDE}&=\E\bigl[Y_{x_2 M_{x_1}}-Y_{x_1 M_{x_1}}\bigr],
\end{align*}
\noindent where $Y_{x_2M_{x_1}}$ is a counterfactual outcome. We can then further expand NIE into the following equations:
\begin{align*}
\text{NIE}&=\E\bigl[Y_{x_2 M_{x_2}}\bigr]-\E\bigl[Y_{x_2 M_{x_1}}\bigr]=\E\Bigl[\E \bigl(Y_{x_2 M_{x_2}}\mid M_{x_2} \bigr) \Bigr] -\E \Bigl[ \E \bigl( Y_{x_2 M_{x_1}} \mid M_{x_1} \bigr) \Bigr] \\ 
&=\E \bigl[ \beta_0+\beta_1M_{x_2}+\beta_2 \mathbbm{1}_{(M_{x_2}>0)}+\beta_3x_2+\beta_4x_2\mathbbm{1}_{(M_{x_2}>0)}+\beta_5x_2M_{x_2} \bigr]\\
&\qquad\qquad -\E \bigl[ \beta_0+\beta_1M_{x_1}+\beta_2 \mathbbm{1}_{(M_{x_1}>0)}+\beta_3x_2+\beta_4x_2\mathbbm{1}_{(M_{x_1}>0)}+\beta_5x_2M_{x_1} \bigr] \\
&=(\beta_1+\beta_5x_2)(\E \bigl[M_{x_2} \bigr]-\E\bigl[M_{x_1}\bigr])+(\beta_2+\beta_4x_2)(\E \bigl[ \mathbbm{1}_{(M_{x_2}>0)} \bigr]-\E \bigl[ \mathbbm{1}_{(M_{x_1}>0)} \bigr])\\
&=\text{NIE}_1+\text{NIE}_2,\\
\text{NIE}_1&=(\beta_1+\beta_5x_2)(\E \bigl[ M_{x_2} \bigr]-\E \bigl[M_{x_1} \bigr]), \\
\text{NIE}_2&=(\beta_2+\beta_4x_2)\bigl(\text{expit}(\gamma_0+\gamma_1 x_1)-\text{expit}(\gamma_0+\gamma_1 x_2)\bigr),
\end{align*}
\noindent where 
\begin{align*}
    &\E\bigl[M_{x_2}\bigr]= \sum_{k=1}^K \psi_k(1-\Delta_{x_2})\mu_{k_{x_2}} + (1-\sum_{k=1}^K \psi_k)(1-\Delta_{x_2})\mu_{K+\Delta_{x_2}}, \\
    &\E\bigl[M_{x_1}\bigr]=\sum_{k=1}^K \psi_k(1-\Delta_{x_1})\mu_{k_{x_1}} + (1-\sum_{k=1}^K \psi_k)(1- \Delta_{x_1})\mu_{K+\Delta_{x_1}}, \\
    & \Delta_{x}=\text{expit}(\gamma_0+\gamma_1x), \\
    &\mu_{k_{x}} = \text{expit}{(\alpha_{0k}+\alpha_{1k}x)},  \hspace{0.2cm} k=1,\dots,K+1,
\end{align*}
\noindent where $\text{expit}(\cdot)$ is the inverse function of $\text{logit}(\cdot)$, $F_{M_x}(m)$ denotes the cumulative distribution function of $M_x$, and $dF_{M_x}(m)$ denotes the Stieltjes integration \citep{stiel} with respect to $F_{M_x}(m)$. The NIE, $\text{NIE}_1$, $\text{NIE}_2$, and NDE can be estimated by substituting the estimated parameters into the corresponding expressions. Confidence intervals (CI) are derived using the multivariate delta method described in the Appendix, while bootstrap procedures provide an alternative means for estimating standard errors and constructing CI \citep{efron1986}. The quantity $\text{NIE}_1$ represents the mediation effect associated with changes in the mediator on its numeric scale, whereas $\text{NIE}_2$ captures the mediation effect attributable to the binary status change of the mediator from zero to a non-zero status. This decomposition is presented in Figure \ref{fig:mediation} below, which displays two indirect causal pathways from $X$ to $Y$ through the mediator $M$. Causal mediation analyses requires assumptions to identify counterfactual outcomes \citep{Imai, VanderWeele2009}. Similar to MarZIC, we make sequential ignorability assumption for our inference.

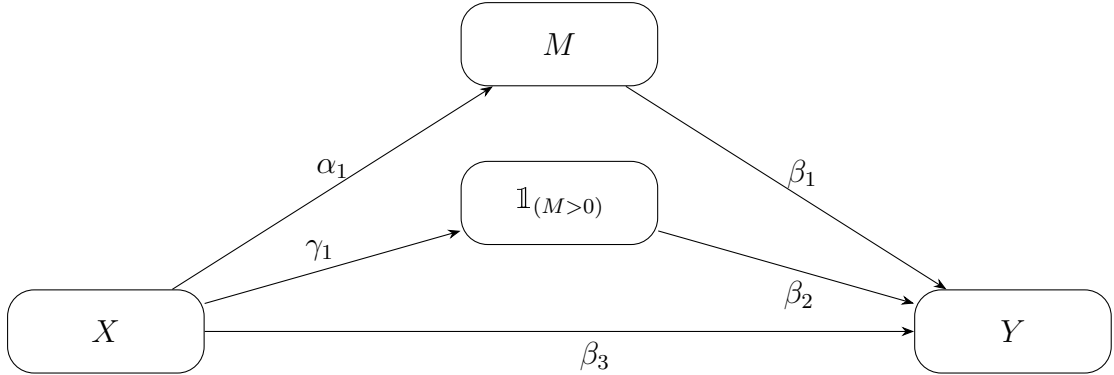
\begin{figure}[htbp]
\centering
\begin{tikzpicture}[
    >=Stealth,
    every node/.style={font=\large},
    box/.style={
        draw,
        rounded corners=10pt,
        minimum width=2.6cm,
        minimum height=1.1cm,
        inner sep=4pt
    }
]

\node[box] (X) at (0,0) {$X$};
\node[box] (M) at (6,3.8) {$M$};
\node[box] (I) at (6,1.7) {$\mathbbm{1}_{(M>0)}$};
\node[box] (Y) at (12,0) {$Y$};

\draw[->] (X) -- node[above,pos=0.5] {$\alpha_1$} (M);
\draw[->] (X) -- node[above,pos=0.45] {$\gamma_1$} (I);
\draw[->] (X) -- node[below,pos=0.55] {$\beta_3$} (Y);
\draw[->] (M) -- node[above,pos=0.55] {$\beta_1$} (Y);
\draw[->] (I) -- node[below,pos=0.55] {$\beta_2$} (Y);

\end{tikzpicture}
\caption{Causal mediation diagram for the ZIBM framework, showing the direct effect of $X$ on $Y$ and indirect effects through $M$ and $\mathbbm{1}_{(M>0)}$}.
\label{fig:mediation}
\end{figure}

\section{Parameter Estimation}\label{sc:estimation}
The indicator variable $c_i$ is defined as follows: $c_i=0$, if the true $m$ is zero; $c_i=k$, if the true $m$ arises from the $k$th Beta distribution. The observed data for each subject can be denoted by the vector $(Y,R,M^*,L,X)$, where $R=\mathbbm{1}_{(M^*>0)}$ and the subject index is suppressed. The log-likelihood function can then be written as follows:
\begin{align*}
    \ell&=\ln\Bigl(\prod_{i=1}^N \prod_{k=0}^{K+1} \bigl(\Psi_{ik}f(y_i,r_i,m_i^* \mid x_i,l_i,c_i=k)\bigr)^{\mathbbm{1}_{(c_i=k)}}\Bigr) \\
    &=\sum_{i=1}^N \sum_{k=0}^{K+1} \mathbbm{1}_{c_i=k} \bigl(\ln(\Psi_{ik}) + \ln(f(y_i,r_i,m_i^* \mid x_i,l_i,c_i=k))\bigr) \\
    &=\sum_{i=1}^N \sum_{k=0}^{K+1} \mathbbm{1}_{c_i=k} \bigl( \ln(\Psi_{ik}) + \ell_{ik} \bigr),
\end{align*}
\noindent where $\Psi_{i0} = \Delta_i$,$\Psi_{ik} = (1-\Delta_i)\psi_{k}, k=1,\dots,K$, and $\Psi_{iK+1}=(1-\Delta_i)(1-\sum_{k=1}^K\psi_k)$. 

Subjects can be partitioned into two groups with group 1 having $m_i^*>0$ and group 2 having $m_i^*=0$. The complete likelihood can then be expressed as: 

$$
\ell = \sum_{i \in \text{group1}}^N \sum_{k=1}^{K+1} \mathbbm{1}_{c_i=k} \bigl(\ln(\Psi_{ik}) + \ell_{ik}^1 \bigr) + 
\sum_{i \in \text{group2}}^N \sum_{k=0}^{K+1} \mathbbm{1}_{c_i=k} \bigl(\ln(\Psi_{ik}) + \ell_{ik}^2 \bigr),
$$

\noindent where $\ell_{ik}^1$ denotes $\ell_{ik}$ for group 1 and $\ell_{ik}^2$ denotes $\ell_{ik}$ for group 2. The complete likelihood is decomposed into three likelihood components.

In group 1, we have $m_i^*=m_i$. Thus, for any $k=1,\dots, k+1,$
\begin{align*}
    \ell_{ik}^1 & = \ln(f(y_i,r_i,m_i^* \mid x_i,l_i,c_i=k)) \\
    &= \ln(f(y_i,r_i \mid m_i^*,x_i,l_i,c_i=k) f(m_i^* \mid x_i,l_i,c_i=k)) \\
    &= \ln(f(y_i \mid m_i^*,x_i,l_i,c_i=k)f(r_i \mid m_i^*,x_i,l_i,c_i=k)f(m_i^* \mid x_i,c_i=k)) \\
    &= \ln(f(y_i \mid m_i^*,x_i)) + \ln(f(m_i^* \mid x_i,c_i=k)) \\
    &=-0.5\ln(2\pi)-\ln(\delta)-\frac{\big(y_i-\beta_0-\beta_1m_i^*-\beta_2-(\beta_3+\beta_4)x_i-\beta_5x_im_i^*\big)^2}{2\delta^2}\\
    & \qquad \qquad +\ln(\mathrm{Beta}(m_i^* \mid \mu_k,\phi)).
\end{align*}

In group 2 where $m_i^*=0$, we have: 
\begin{align*}
    \ell_{i0}^2 &= \ln(f(y_i,r_i=0,m_i^*=0 \mid x_i,l_i,c_i=0))= \ln(f(y_i,r_i=0 \mid x_i,l_i,c_i=0)) \\
    &= \ln(f(y_i \mid x_i,l_i,c_i=0)f(r_i=0 \mid x_i,l_i,c_i=0)) \\
    &= \ln(f(y_i \mid x_i,l_i,c_i=0)) \\
    &=-0.5\ln(2\pi)-\ln(\delta) -\frac{(y_i-\beta_0-\beta_3x_i)^2}{2\delta^2},
\end{align*}

And for any $k = 1, \dots, k+1,$
\begin{align*}
    \ell_{ik}^2 &= \ln(f(y_i,r_i=0,m_i^*=0 \mid x_i,l_i,c_i=k)) = \ln(f(y_i,r_i=0 \mid x_i,l_i,c_i=k)) \\
    &= \ln(\int\limits_0^{1/l_i} f(y_i,r_i=0 \mid x_i,l_i,c_i=k,m)dF(m \mid x_i,l_i,c_i=k)) \\
    &= \ln(\int\limits_0^{1/l_i} f(y_i \mid x_i,m)dF(m \mid x_i,l_i,c_i=k)) \\
    &= \ln(\int\limits_0^{1/l_i} f(y_i \mid x_i,m)f(m \mid x_i,l_i,c_i=k))dm \\
    &= -0.5\ln(2\pi)-\ln (\delta)+\ln(\int\limits_0^{1/l_i} h(y_i \mid x_i,m)\mathrm{Beta}(m \mid \mu_k,\phi))dm , 
\end{align*}
\noindent where $$
h(y_i \mid x_i,m)=\exp\biggl(-\frac{\bigl(y_i-\beta_0-\beta_1m-\beta_2-(\beta_3+\beta_4)x_i-\beta_5x_im\bigr)^2}{2\delta^2}\biggr).$$

\section{EM Algorithm}\label{sc:EM}
The Expectation-Maximization (EM) algorithm \citep{EM.original} is used to obtain the maximum likelihood estimates of parameters. Throughout this iterative estimation procedure, we let $\Theta$ denote the full parameter vector containing all parameters and let $\Theta^0$ denote the initial parameter values. At each iteration, the conditional expectation of the complete-data log-likelihood is computed given the observed data.
\subsection{E-Step}
The conditional expectation of the log-likelihood function is defined as follows:
\begin{align*}
    Q(\Theta \mid \Theta^0) = \E \bigl[ \ell \mid m_i^*,y_i,r_i,l_i,\Theta^0 \bigr]  &= \sum_{i \in \text{group1}}^N \sum_{k=1}^{K+1} \E \bigl[ \mathbbm{1}_{c_i=k} \mid m_i^*,y_i,r_i=1,l_i,\Theta^0 \bigr] (\ln(\Psi_{ik}) + \ell_{ik}^1) \\
    & \qquad\qquad + \sum_{i \in \text{group2}}^N \sum_{k=0}^{K+1} \E\bigl[\mathbbm{1}_{c_i=k} \mid m_i^*,y_i,r_i=0,l_i,\Theta^0\bigr](\ln(\Psi_{ik}) + \ell_{ik}^2).
\end{align*}

\noindent Let $\tau_{ik}^1(\Theta^0) = \E \bigl[ \mathbbm{1}_{c_i=k} \mid m_i^{*},y_i,r_i=1,l_i,\Theta^0 \bigl]$ and $\tau_{ik}^2(\Theta^0) =\E \bigl[ \mathbbm{1}_{c_i=k} \mid m_i^*,y_i,r_i=0,l_i,\Theta^0 \bigr]$. 

\begin{align*}
    \tau_{ik}^1(\Theta^0) &= \E \bigl[ \mathbbm{1}_{c_i=k} \mid m_i^*,y_i,r_i=1,l_i,x_i,\Theta^0 \bigr] = P \bigl(c_i=k \mid m_i^*,y_i,r_i=1,l_i,x_i,\Theta^0 \bigr) \\
    &= \frac{f(m_i^*,y_i,r_i=1 \mid \Theta^0,c_i=k,l_i,x_i)\,P(c_i=k \mid \Theta^0)}{\sum_{k=1}^{K+1} f(m_i^*,y_i,r_i=1 \mid \Theta^0,c_i=k,l_i,x_i)\, P(c_i=k \mid \Theta^0)} \\
    &= \frac{\left. \Psi_{ik}\exp(\ell_{ik}^1)\right|_{\Theta=\Theta^0}}{\left.\sum_{k=1}^{K+1}\Psi_{ik}\exp(\ell_{ik}^1)\right|_{\Theta=\Theta^0}}
\end{align*}

\begin{align*}
    \tau_{ik}^2(\Theta^0) &= \E\bigl[\mathbbm{1}_{c_i=k} \mid m_i^*=0,y_i,r_i=0,l_i,x_i,\Theta^0 \bigr] = P \bigl(c_i=k \mid m_i^*=0,y_i,r_i=0,l_i,x_i,\Theta^0 \bigr) \\
    &= \frac{f(y_i,r_i=0 \mid \Theta^0,c_i=k,l_i,x_i) \, P(c_i=k \mid \Theta^0)}{\sum_{k=0}^{K+1} f(y_i,r_i=0 \mid \Theta^0,c_i=k,l_i,x_i) \, P(c_i=k \mid \Theta^0)} \\
    &=\frac{\left. \Psi_{ik}\exp(\ell_{ik}^2)\right|_{\Theta=\Theta^0}}{\left. \sum_{k=0}^{K+1}\Psi_{ik}\exp(\ell_{ik}^2)\right|_{\Theta=\Theta^0}}
\end{align*}

\noindent Thus, we have: 
\begin{align*}
    Q(\Theta \mid \Theta^0)=\E \bigl[ \ell \mid m_i^*,y_i,r_i,l_i,\Theta^0 \bigr] &=\sum_{i \in \text{group1}}^N \sum_{k=1}^{k+1} \tau_{ik}^1(\Theta^0) \bigl(\ln(\Psi_{ik}) + \ell_{ik}^1\bigr) \\
    & + \sum_{i \in \text{group2}}^N \sum_{k=0}^{k+1} \tau_{ik}^2(\Theta^0) \bigl(\ln(\Psi_{ik}) + \ell_{ik}^2 \bigr).
\end{align*}

\subsection{M-Step}
In the M-step, $Q(\Theta \mid \Theta^0)$ is maximized with respect to $\Theta$ to obtain updated parameter estimate denoted as $\Theta^1$. The E-step and M-step are repeated until the Euclidean distance between successive parameter estimates $\lvert\Theta^0 - \Theta^1\rvert <10^{-8}$. The final estimate $\Theta^1$ is taken as the maximum likelihood estimator (MLE).

\subsection{Information Matrix}
Let $\Theta^*$ denote the MLE. The information matrix is computed as below which can be used to obtain standard errors \citep{oakes}.
$$
I= - \left. \frac{\partial Q(\Theta \mid \Theta^*)}{\partial \Theta^2} \right|_{\Theta=\Theta^*} - 
\left. \frac{\partial Q(\Theta \mid \Theta^0)}{\partial \Theta \partial \Theta^0}\right|_{\Theta=\Theta^*, \Theta^0=\Theta^*}
$$

\section{Simulation}\label{sc:simulations}
Simulation studies were conducted to assess the performance of the ZIBM model. Two settings were considered. In the first setting, data were generated from a univariate beta mixture distribution to evaluate the proposed model under correct model specification. The performance of the ZIBM model was compared with the nonparametric causal mediation approach of Imai, Keele, and Tingley \citep{Imai} (hereafter referred to as the IKT approach). Although the IKT method adopts a similar definition of the natural indirect effect (NIE) compared to ours, it was not developed for microbiome data and was expected to perform poorly in the presence of zero inflation. In the second setting, the ZIBM was applied to simulated microbiome data with many taxa. The objective was to identify significant mediators, and the results were compared with those obtained from CCMM \citep{Sohn2019}, which was specifically developed for microbiome data.

In both settings, the exposure $X$ was generated from a Bernoulli distribution, $\mathrm{Ber}(0.5)$. False zeros were introduced based on the LOD mechanism in Equation (\ref{eq:RAzero1}) of Subsection \ref{sc:mechanismzero}. The library sizes were randomly sampled from a real study \citep{Arthur2013} (see Section \ref{sc:realdata} for more details), ranging from 31,607 to 911,652. In each setting, 100 datasets were generated to evaluate model performance. Asymptotic variances were used to compute coverage probabilities for each parameter, and the multivariate delta method was employed to derive standard errors for the mediation effects.

\subsection{Simulation Setting I}
In this setting, the mediator $M$ was generated from a univariate beta mixture distribution using Equations (\ref{ge:2}) to (\ref{ge:23}). The outcome $Y$ was generated from Equation (\ref{ge:1}). Two simulation scenarios were considered with different numbers of beta mixture components for the mediator. The first scenario had three beta mixture components with 300 samples in each of the 100 datasets. In this scenario, $41.3\%$ of the observed $M$ values were zero, among which $44.6\%$ were false zeros generated from the LOD mechanism. The second scenario had two beta mixture components with 200 samples in each of the 100 datasets. In this scenario, $51.2\%$ of the observed $M$ values were zero, among which $75.6\%$ were false zeros. The results were shown in Table \ref{table:sim1}. From the results, we can see that both scenarios had accurate estimates of the unknown parameters and mediation effects. The bias of all the parameters was small. Some parameters had higher bias percentage due to small true values. Note that the bias percentage of $\alpha_{01}$ in the three-component beta mixture model is not available since the true value is 0. The coverage probability (CP) from the $95\%$ CIs was close to $95\%$. On the other hand, IKT approach had relatively poor performance with respect to both CP and bias compared to ZIBM. The bias was likely due to two factors: (i) failure to account for false zeros and (ii) the lack of NIE decomposition into $\text{NIE}_1$ and $\text{NIE}_2$.

\begin{table}[ht!]
\caption{Simulation results for mediators generated from three- and two-beta mixture distributions. The mean of estimates, bias, percentage of bias, mean of asymptotic standard error, and coverage probability from $95\%$ CIs for mediation effects and each parameter are reported. The mediation effect from the IKT approach was also reported at the bottom.}
\label{table:sim1}
\centering
\resizebox{\linewidth}{!}{
\begin{tabular}{ccccccccccccccc}
  \hline
  & \multicolumn{6}{c}{Three-beta mixture distribution} && \multicolumn{6}{c}{Two-beta mixture distribution} \\
  \cline{2-7} \cline{9-14}
   Parameter & True & Mean Estimate & Bias & Bias(\%) & Mean SE & CP(\%) && True & Mean Estimate & Bias & Bias(\%) & Mean SE & CP(\%) \\ 
  \hline 
  \multicolumn{14}{c}{\textbf{ZIBM}}\\
  NIE$_1$ & -0.28 & -0.27 & 0.01 & -2.23 & 0.17 & 92 && -0.18 & -0.18 & -0.00 & 0.75 & 0.16 & 97 \\  
  NIE$_2$ & 0.57 & 0.53 & -0.04 & -7.54 & 0.36 & 94 && 0.57 & 0.60 & 0.03 & 5.65 & 0.46 & 96 \\ 
  NIE & 0.29 & 0.25 & -0.04 & -12.66 & 0.45 & 95 && 0.39 & 0.42 & 0.03 & 7.87 & 0.58 & 95 \\  
  $\beta_0$ & -5.00 & -5.01 & -0.01 & 0.22 & 0.20 & 92 && -5.00 & -4.97 & 0.03 & -0.55 & 0.23 & 95 \\
  $\beta_1$ & 10.00 & 9.94 & -0.06 & -0.56 & 0.45 & 89 && 10.00 & 9.96 & -0.04 & -0.39 & 0.32 & 99 \\
  $\beta_2$ & 8.00 & 8.03 & 0.03 & 0.39 & 0.23 & 94 && 8.00 & 7.97 & -0.03 & -0.36 & 0.27 & 94 \\  
  $\beta_3$ & 1.00 & 1.16 & 0.16 & 15.68 & 0.31 & 92 && 1.00 & 0.93 & -0.07 & -6.92 & 0.38 & 91 \\  
  $\beta_4$ & 1.00 & 0.80 & -0.20 & -20.09 & 0.35 & 95 && 1.00 & 1.09 & 0.09 & 8.60 & 0.42 & 90 \\ 
  $\beta_5$ & 1.00 & 1.08 & 0.08 & 8.48 & 0.71 & 94 && 1.00 & 1.02 & 0.02 & 1.94 & 0.49 & 94 \\ 
  $\gamma_0$ & -1.50 & -1.52 & -0.02 & 1.11 & 0.21 & 98 && -1.50 & -1.50 & 0.00 & -0.14 & 0.26 & 98 \\  
  $\gamma_1$ & -0.50 & -0.45 & 0.05 & -9.02 & 0.33 & 93 && -0.50 & -0.54 & -0.04 & 7.62 & 0.41 & 96 \\  
  $\phi$ & 10.00 & 11.05 & 1.05 & 10.49 & 2.48 & 94 && 10.00 & 10.66 & 0.66 & 6.55 & 1.88 & 94 \\  
  $\delta$ & 1.00 & 1.02 & 0.02 & 2.17 & 0.04 & 92 && 1.00 & 0.99 & -0.01 & -1.18 & 0.05 & 94 \\  
  $\alpha_{01}$ & 0.00 & 0.00 & 0.00 & - & 0.16 & 93 && 1.00 & 1.02 & 0.02 & 2.28 & 0.14 & 94 \\ 
  $\alpha_{02}$ & -2.00 & -2.01 & -0.01 & 0.69 & 0.26 & 96 && -5.00 & -5.04 & -0.04 & 0.72 & 0.25 & 95 \\  
  $\alpha_{03}$ & -5.00 & -5.11 & -0.11 & 2.10 & 0.32 & 96 & & - & - & - & - & - & - \\ 
  $\alpha_{11}$ & -0.50 & -0.48 & 0.02 & -3.60 & 0.22 & 92 && -0.50 & -0.52 & -0.02 & 4.62 & 0.18 & 93 \\  
  $\alpha_{12}$ & -0.50 & -0.49 & 0.01 & -2.75 & 0.37 & 95 && -0.50 & -0.51 & -0.01 & 1.94 & 0.31 & 95 \\  
  $\alpha_{13}$ & -0.50 & -0.45 & 0.05 & -10.72 & 0.51 & 92 && - & - & - & - & - & -\\ 
  $\psi_1$ & 0.20 & 0.20 & 0.00 & 1.73 & 0.03 & 96 && 0.30 & 0.30 & 0.00 & 0.03 & 0.04 & 95 \\  
  $\psi_2$ & 0.30 & 0.30 & -0.00 & -0.69 & 0.05 & 91 && - & - & - & - & - & - \\ 
  \multicolumn{14}{c}{\textbf{IKT}}\\
  NIE & 0.29 & -0.26 & -0.55 & -190.35 & 0.23 & 29 && 0.39 & -0.14 & -0.53 & -136.54 & 0.47 & 79 \\ 
   \hline
\end{tabular}}
\end{table}

\subsection{Simulation Setting II}
In this simulation setting, we proposed an approach to generate microbiome RA data from a Dirichlet distribution, with only the first two taxa correlated with the covariate $X$. Therefore, based on the definition of mediation effect in Subsection \ref{NIE_NDE}, only the first two taxa have non-zero NIEs. The steps to generate RA for each taxon with $K+1$ taxa are provided in Appendix \ref{appen_2}.

The outcome $Y$ model was generated as follows: 
\begin{align*}
    Y=\beta_0+\beta_1M_1+\beta_2\mathbbm{1}_{M_1>0}+\beta_3X+\beta_4X\mathbbm{1}_{M_1>0}+\beta_5XM_1+\epsilon
\end{align*}
where $M_1$ denotes the first taxon. The outcome $Y$ was marginally associated with the first taxon, and due to the compositional structure, outcome $Y$ was also marginally associated with other taxa. $(\beta_0,\beta_1,\beta_2,\beta_3,\beta_4,\beta_5)=(4,100,2,1,1,1)$ and $\epsilon$ follows a standard normal distribution. We set $\gamma_0=0$ and $\gamma_1=-3$ for the first taxon, and for other taxon $\gamma_0$ was adjusted to make their zero percentage close to $50\%$. $\gamma_1$ for the other taxon was set to 0 to make sure their $\text{NIE}_2$ was 0. The $\boldsymbol{\alpha}$ was adjusted to ensure that approximately $40\%$ of all zeros in the first taxon were false zeros.

Model performance was evaluated using three metrics: Recall, Precision, and F1, defined as follows:
\begin{align*}
    \text{Recall}=\frac{TP}{TP+FN},\hspace{0.5cm} \text{Precision}=\frac{TP}{TP+FP},\hspace{0.5cm} \text{F1}=\frac{2}{\frac{1}{\text{Recall}}+\frac{1}{\text{precision}}}
\end{align*}
where TP, TN, FP, FN are true positive, true positive, false positive, and false negative, respectively. Multiple testing in our model was adjusted using the Benjamini-Hochberg procedure \citep{BHadjust} with false discovery rate (FDR) set to $20\%$. Accordingly, the target precision rate was $80\%$.

The results for simulation setting II are presented in Table \ref{table:sim2}. CCMM, which was specifically developed for microbiome datasets, was included for comparison. The proposed method achieved substantially better performance, reaching the target precision rate with a reasonable recall, whereas CCMM exhibited lower recall and remarkably lower precision. In the 300 taxon case, CCMM failed to produce results due to computational challenge.

\begin{table}[ht]
\caption{Simulation results for the simulation setting II with comparison to CCMM. }
\label{table:sim2}
\centering
\resizebox{\linewidth}{!}{
\begin{tabular}{ccccccccccccc}
  \hline
  && \multicolumn{3}{c}{Recall(\%)} && \multicolumn{3}{c}{Precision(\%)} && \multicolumn{3}{c}{F1(\%)} \\
  \cline{3-5} \cline{7-9} \cline{11-13} \noalign{\vskip .06in}
$K+1$ &n &ZIBM & ZIBM & CCMM && ZIBM & ZIBM & CCMM && ZIBM & ZIBM & CCMM \\ 
&& (NIE$_1$)&(NIE$_2$)&&&(NIE$_1$)&(NIE$_2$)&&&(NIE$_1$)&(NIE$_2$)& \\
  \hline
10 & 200 & 80.00 & 84.80 & 70.50 && 97.70 & 83.50 & 19.60 && 84.90 & 77.10 & 30.50 \\  
  10 & 300 & 82.50 & 86.00 & 72.00 && 93.30 & 87.50 & 17.60 && 84.10 & 79.00 & 28.20 \\ 
  25 & 200 & 87.50 & 85.00 & 69.00 && 80.40 & 90.80 & 16.40 && 80.20 & 80.10 & 26.20 \\ 
  25 & 300 & 98.00 & 86.00 & 68.50 && 91.40 & 92.80 & 14.10 && 93.30 & 80.70 & 23.30 \\ 
  50 & 200 & 97.50 & 91.90 & 67.50 && 99.30 & 95.30 & 14.60 && 97.90 & 88.60 & 23.70 \\ 
  50 & 300 & 88.80 & 98.00 & 68.00 && 96.60 & 87.80 & 12.40 && 90.50 & 90.30 & 20.80 \\ 
  100 & 200 & 83.50 & 95.00 & 79.00 && 89.90 & 91.70 & 17.30 && 83.20 & 89.30 & 27.70 \\ 
  100 & 300 & 90.90 & 96.00 & 83.00 && 85.80 & 90.40 & 14.20 && 84.90 & 90.10 & 24.00 \\ 
  300 & 200 & 97.00 & 96.00 & - && 94.70 & 83.80 & - && 94.70 & 85.10 &-  \\ 
  300 & 300 & 99.50 & 96.80 &-  && 99.30 & 91.10 &-  && 99.20 & 89.70 &-  \\ 
   \hline
\end{tabular}}
\end{table}

\section{Real Data Application}\label{sc:realdata}
\begin{figure}[ht]
  \begin{center}
  \centerline{\includegraphics[scale=0.9]{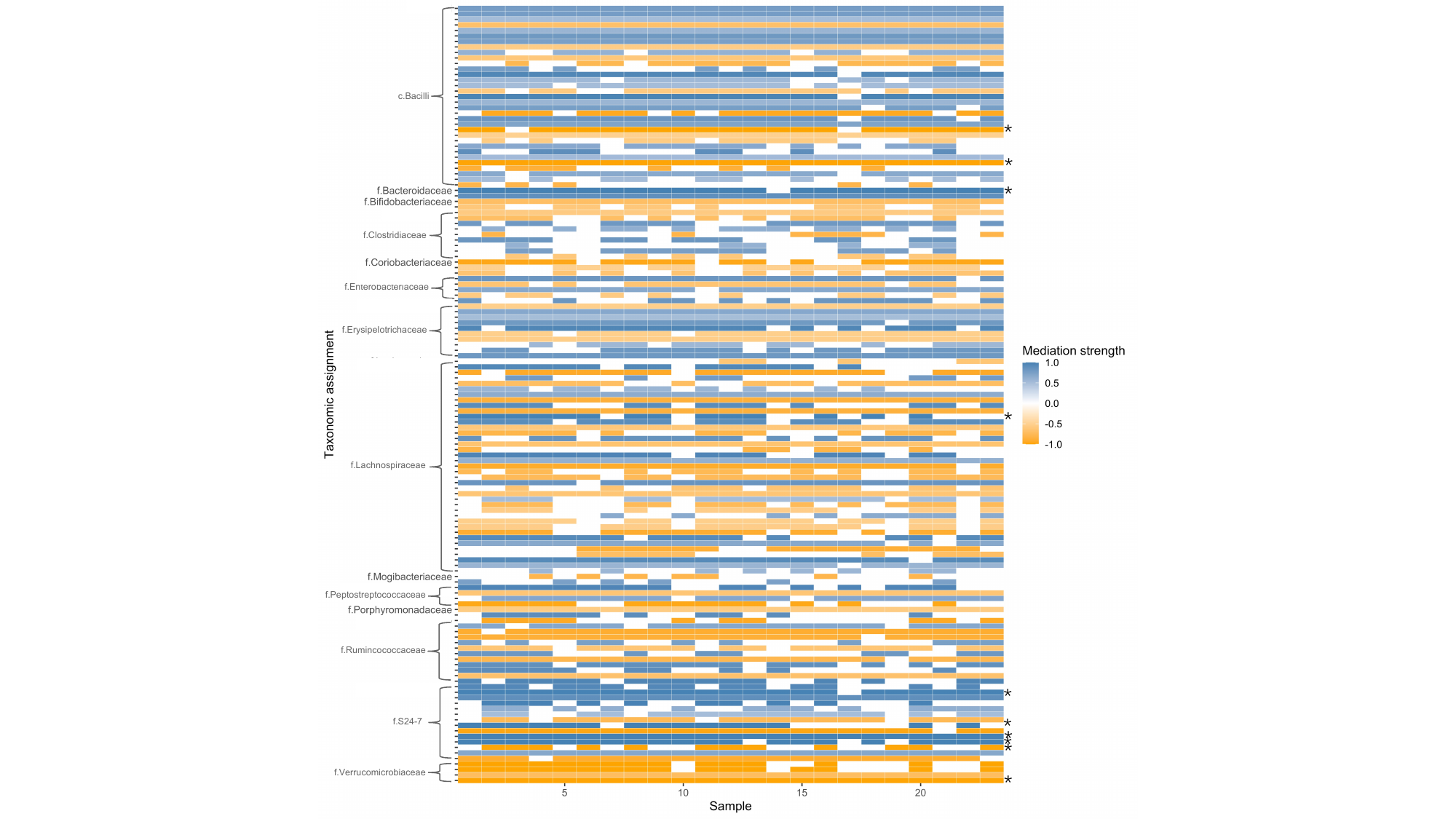}}
  \end{center}
  \caption{Heatmap of mediation strength based on $\text{NIE}_1$ in VSL\#3 study. The mediation strength is measured by $(1-p)$, where $p$ is the unadjusted p-value. A negative sign indicates a negative $\text{NIE}_1$. Taxonomic assignments are labeled on the vertical axis. Samples are labeled on the horizontal axis. Absence of an OTU in a sample is shown as blank in the heatmap.}
  \label{VSL_heatmap_NIE1}
\end{figure}

In this section, we applied our proposed method to a study conducted by Arthur et al. \citep{Arthur2013} to investigate the effect of VSL\#3 on inflammation-associated CRC in mice after the onset of inflammation. The VSL\#3 is a cocktail (Sigma-Tau Pharmaceuticals, Inc.) of eight strains of lactic acid-producing bacteria: \textit{Lactobacillus plantarum, Lactobacillus delbrueckii subsp. Bulgaricus, Lactobacillus paracasei, Lactobacillus acidophilus, Bifidobacterium breve, Bifidobac- terium longum, Bifidobacterium infantis}, and \textit{Streptococcus salivarius subsp}. Recent studies has shown that orally administered VSL\#3 reduced intestinal inflammation in humans \citep{Gionchetti2000,Sood2009,Madsen2001,Pagnini2010}. The 24-week study used 24 mice, with 10 in the oral treatment group and 14 in the control group. The microbiome data were obtained from 16S rRNA sequencing with samples collected from stool at the end of the study. For each operational taxonomic unit (OTU) obtained from the sample, RA was analyzed as a mediator using our model. The outcome variable was dysplasia score, an ordinal categorical variable measuring the abnormality of cell growth, with higher scores indicating greater dysplasia. \cite{Arthur2013} found that depletion of the \textit{Clostridium} bacterial group was associated with VSL\#3 administration and tumorigenesis using mediation by linear models analysis. Applying MarZIC, \cite{MarZIC} identified two significant mediators (i.e., family \textit{S24-7} and class \textit{Bacilli}). The ZIBM model proposed in this study identified 10 significant mediators, where two belonged to the \textit{Bacilli} class, one from the \textit{Bacteroidaceae} family, one from the \textit{Lachnospiraceae} family, one from \textit{Verrucomicrobiaceae} family, and five from the \textit{S24-7} family. See Figure \ref{VSL_heatmap_NIE1} for the heatmap of these results.

The literature supports these findings. Notable \textit{Bacilli}-class members implicated in CRC \citep{gagniere, karpinski}. \textit{Bacteroidaceae} has been found to be significantly enriched in stool samples from patients with inflammatory bowel disease and CRC \citep{dan}. \textit{Lachnospiraceae} has a complex and relatively well-studied association with gut inflammatory disease and CRC. The literature suggests that specific genera of bacteria within the family are protective commensal intestinal symbionts while others are pathogenic \citep{zhang, ma, shah, tang}. \textit{Verrucomicrobiaceae} is not as well-characterized. However, it has been shown to have a context-dependent role in CRC pathogenesis. In inflammatory disease contexts, one of its members, \textit{Akkermansia muciniphila}, acts as a pro-tumoigenic bacteria whereas it is enriched in sporadic CRC tissue and one of the main differentiating phyla in cancer patients compared to healthy controls \citep{soheilipour}. The \textit{S24-7} family has been associated with murine colitis, however, the clinical relevance in humans is yet unknown \citep{lagkouvardos}.

\section{Discussion}\label{sc:discussion}
In this paper, we propose a ZIBM model for mediation analysis with compositional microbiome mediators. The model extends the MarZIC framework by introducing a mixture representation for the non-zero component, which accounts for excess zeros and latent heterogeneity in microbiome RA data. It is formulated within the potential-outcomes framework and provides interpretable estimates of microbiome-mediated causal effects.

Simulation results show that the proposed method performs well under conditions commonly observed in microbiome data, including zero inflation, compositionality, and latent heterogeneity. Compared with existing methods (i.e., IKT \citep{Imai} and CCMM \citep{Sohn2019}), the ZIBM model yields more reliable estimates when false zeros are present or when standard distributional assumptions are violated. The real data analysis demonstrates the utility of the proposed model for microbiome studies with sparse and compositional mediators. The identified mediators are associated with disease-related outcomes while accounting for sparsity and compositional constraints. Compared to the MarZIC method's application of the \cite{Arthur2013} study, the ZIBM framework identified eight additional bacterial families as causal mediators \citep{MarZIC}. All of the bacterial classes and families identified have been associated with high-risk inflammatory diseases and CRC. However, the associations are complex with different genera within one family having different commensal or pathogenic effects. Therefore, further research with specific targeted genera and species using the ZIBM framework is warranted. Our findings suggest that ZIBM may be a more sensitive approach that allows for greater capture of potential causal mediators. Identifying causal mediators that explain how or why an exposure affects an outcome will help physicians, researchers, and policymakers better understand oncogenesis, discover risk factors, and develop interventions.

The proposed model offers several advantages. The zero-inflated beta mixture formulation accounts for both structural and sampling zeros and captures heterogeneity in non-zero RAs. It provides an interpretable mediation framework under the potential-outcomes paradigm for sparse and compositional microbiome data. The ZIBM framework also has several limitations. The number of mixture components need to be specified in advance, which may affect model performance, and computational burden may increase in high-dimensional microbiome settings. Future work includes data-driven selection of mixture components. In summary, the ZIBM model provides a practical approach for mediation analysis with zero-inflated compositional microbiome mediators.

\section{Appendix}

\subsection{Data Generation Process} \label{appen_2}
Simulated data are adapted from the MarZIC \citep{MarZIC} framework, with modifications described below. Let $Q+1$ be the number of taxa, for the $i$th subject, the microbiome data generation steps are listed below: \newline

\noindent \textit{\textbf{Step 1:}} Decide which component of the $k$th component mixture will be used for this subject. For illustration purpose, $k$ is set to be 2. In the simulation, the weights of the 2-component mixture are $\psi_1=0.3$ and $\psi_2=0.7$. Generate a Bernoulli random variable from $\mathrm{Ber}(0.3)$. If $\mathrm{Ber}(0.3) = 1$, the first component is used. Otherwise, the second component is used. 

\noindent \textit{\textbf{Step 2:}} Generate true zeros for all taxa. The zeros for a taxon were generated using a $\mathrm{Ber}(\Delta)$. So the probability of the taxon being 0 is equal to $\Delta$. For taxon 1, we set $\Delta=0$ so that there is no zero in taxon 1. For taxon 2, $\Delta$ was calculated using the selected mixture component from Step 1. For all other taxa, $\gamma_{0}$ were generated $\mathrm{Unif}(1,2)$ and $\gamma_1=0$ for $\Delta$. Thus, only the absence (or presence) of taxon associated with $X$. The total percentage of zeros was between 68.8$\%$ and 81.6$\%$ with $K+1$ ranging from 10 to 500, which indicates high data sparsity.

\noindent \textit{\textbf{Step 3:}} Generate RA for the non-zero taxa from a Dirichlet distribution. Assume we had $P$ non-zero taxa (from Step 1) indexted by $(t_1, t_2, \dots, t_p)$ in the ascending order meaning $t_1 < \cdots < t_p$. Here $t_1=1$ since the first taxon does not have any zeros from Step 1. The RA of those non-zero taxa was generated by the $P$-dimensional Dirichlet distribution with the dispersion parameter $\phi$ and mean parameter $(\mu_{t_1}, \cdots, \mu_{t_p})$ that satisfies $\sum_{p=1}^p \mu_{t_p} = 1$. The dispersion parameter $\phi$ was set to be 10 to mimic the overdispersion in real data. The values of mean parameters were chosen in a way such that it has smaller values for taxa with larger $\Delta$'s in Step 1 so that taxa with lower abundance are more likely to have zeros. More specifically, the mean parameters were determined as follows:
If $t_2=2$ (i.e., taxa 2 is one of the non-zero taxa):
\begin{align*}
&\mu_{t_1}=\frac{\exp{(\alpha_0^1)}}{\sum_{q=1}^{Q+1}\exp{(\alpha_0^q)}} \times \frac{1}{1+\exp{(\alpha_{0}+\alpha_{1}X)}}, \\
&\mu_{t_2}= \frac{\exp{(\alpha_0^1)}}{\sum_{q=1}^{Q+1}\exp{(\alpha_0^q)}} \times \frac{\exp{(\alpha_{0}+\alpha_{1}X)}}{1+\exp{(\alpha_{0}+\alpha_{1}X)}}, \\
&\mu_{t_p}= \Biggl(1- \frac{\exp{(\alpha_0^1)}}{\sum_{q=1}^{Q+1}\exp{(\alpha_0^q)}} \Biggr)\times \frac{\exp{(\alpha_{0}^{t_{p}})}}{\sum_{p=3}^P\exp{(\alpha_{0}^{t_{p}})}}, p \in \{3, \dots , P \} ,
\end{align*}
where $\alpha_0 = -2$, $\alpha_{1}=5$, $\alpha_{0}^{1}$ was set to be the value such that the false zeros for taxon 2 that will be generated in next step will be around $20\%$, and $\alpha_{0}^{k}$, $k \in \{2, \dots, K+1\}$ were generared from $\mathrm{Unif}(0,1)$. If $t_2 > 2$ (i.e., taxa 2 is not one of the non-zero taxa): 
\begin{align*}
&\mu_{t_1}=\frac{\exp{(\alpha_0^1)}}{\sum_{q=1}^{Q+1}\exp{(\alpha_0^q)}}, \\
&\mu_{t_p}= \Biggl(1- \frac{\exp{(\alpha_0^1)}}{\sum_{q=1}^{Q+1}\exp{(\alpha_0^q)}} \Biggr)\times \frac{\exp{(\alpha_{0}^{t_{p}})}}{\sum_{p=2}^P\exp{(\alpha_{0}^{t_{p}})}}, p \in \{2, \dots , P \} ,
\end{align*}
Under this data generation, only the RA of taxa 1 and taxa 2 were dependent on $X$. The RA of taxa 3 to taxa $K$ were independent of $X$.

\noindent \textit{\textbf{Step 4:}} Generate sample absolute abundance (SAA) and false zeros from a multinomial distribution. Let $(\mathcal{R}_{t_1}, \dots, \mathcal{R}_{t_p}$ denote the RA generated in Step 2 for the P non-zero taxa thus $\sum_{p=1}^P \mathcal{R}_{t_p}=1$. The $P$-dimensional multinomial distribution with the parameter vector $(\mathcal{R}_{t_1}, \dots, \mathcal{R}_{t_p}$ and the library size (randomly selected from the real data) was used to generate SAA for all the P non-zero taxa. Those taxa with SAA=0 generated from the multinomial distribution are false zeros.

\noindent \textit{\textbf{Step 5:}} Getting final RA for all non-zero taxa. After SAA were generated for all non-zero taxa in Step 3, the SAA were divided by the library size to get the final RA for all non-zero taxa.

\noindent \textit{\textbf{Step 6:}} Repeat the above Steps 1-5 for each subject to get a full data set of microbiome data for 200 subject.

\section*{Funding}
\vspace{-0.5cm}
The first author (S.A.) was supported in part by National Cancer Institute Cancer Center Support Grant P30CA196521 awarded to the Tisch Cancer Center of the Icahn School of Medicine at Mount Sinai. The funder had no role in study design, data collection and analysis, decision to publish, or preparation of the manuscript.

\vspace{0.1cm}
\bibliography{JabRef.bib}
\bibliographystyle{apalike}

\end{document}